\documentclass[fleqn, usenatbib]{mnras}
\usepackage{graphicx}
\usepackage{newtxtext,newtxmath}
\usepackage[T1]{fontenc}
\usepackage{xcolor}
\usepackage[utf8]{inputenc}
\usepackage{CJKutf8}
\usepackage{amsmath}
\usepackage{multirow}

\author[W. Zhang et al.]{
Wenda Zhang~
\begin{CJK*}{UTF8}{gbsn}
(张文达)
\end{CJK*}
$^{1,2}$\thanks{wdzhang@nao.cas.cn},
Michal Dov\v{c}iak$^2$,
Michal Bursa$^2$,
Ji\v r\'i Svoboda$^2$,
Vladim\'ir Karas$^2$
\\
$^1$National Astronomical Observatories, Chinese Academy of Sciences, A20 Datun Road, Beijing 100101, China\\
$^2$Astronomical Institute of the Czech Academy of Sciences, Bo{\v c}n{\'i} II 1401, CZ-14100 Prague, Czech Republic}

\title[Emissivity profile]{Inferring the iron K emissivity profiles of accretion discs irradiated by extended coronae}

\date{Accepted XXX. Received YYY; in original form ZZZ}

\pubyear{2023}

\begin{document}
\label{firstpage}
\pagerange{\pageref{firstpage}--\pageref{lastpage}}
\maketitle

\begin{abstract}
One of the most promising methods to measure the spin of an accreting black hole is fitting the broad
iron K$\alpha$ line in the X-ray spectrum.
The line profile also depends on the geometry of the hard X-ray emitting corona.
To put constraints on the black hole spin and corona geometry, it is essential to understand how do they
affect the iron K$\alpha$ line emissivity profile.
In this work, we present calculations of the illumination and the iron K$\alpha$
emissivity profiles performed with the Monte-Carlo GR radiative transfer code \textsc{Monk}.
We focus on distinction between the illumination and emissivity profiles, which is in most previous studies neglected. We show that especially for the case of black hole X-ray binaries (BHXRBs), the difference is very large.
For active galactic nuclei (AGNs), the emissivity profile has a more similar shape as the illumination profile, but it is notably steeper in the innermost region within a few gravitational radii.
We find out that the different behavior between AGN and black hole X-ray binary discs is due to the different energy spectra of
the illuminating radiation.
This suggests that the emissivity profile of the iron K$\alpha$ line 
cannot be determined by black hole spin and corona geometry alone and the energy spectrum of the illuminating
radiation has to be taken into account.
We also examined the effect of including the self-irradiation, and find it to be more important than the corona emission 
in BHXRBs.
\end{abstract}


\section{Introduction}
X-ray spectra of Active Galactic Nuclei (AGNs) and black hole X-ray binaries (BHXRBs) exhibit signatures of X-ray reflection,
including the iron K fluorescent line and the Compton hump \citep[see the review of][]{miller_relativistic_2007}.
It is generally believed that they are due to the hard X-ray radiation of the corona illuminating the underlying accretion disc
\citep[see][and references therein]{fabian_x-ray_2010}. The iron line is intrinsically narrow, but 
the observer at infinity sees a broad and asymmetric line profile, due to the
gravitational redshift by the strong gravitational field of the black hole and due to the Doppler broadening
\citep{fabian_x-ray_1989,laor_line_1991}.
The shape of the broad iron line depends sensitively on the black hole spin, and thus
fitting the iron line profile is one of the most promising methods of 
measuring the spin of accreting black holes in both XRBs
\citep[e.g.][]{blum_measuring_2009,miller_stellar-mass_2009,miller_nustar_2013,parker_nustar_2015,el-batal_nustar_2016,
miller_nicer_2018,xu_reflection_2018,xu_hard_2018}
and AGNs \citep[e.g.][]{brenneman_constraining_2006,miniutti_intermediate_2009,schmoll_constraining_2009,fabian_broad_2009,
brenneman_spin_2011,emmanoulopoulos_xmm-newton_2011,reynolds_monte_2012,tan_possible_2012,risaliti_rapidly_2013,
ricci_suzaku_2014,agis-gonzalez_black_2014,sun_multi-epoch_2018,jiang_relativistic_2019}.

To put narrow constraints on black hole spins it is important to know the radial emissivity profile of the iron fluorescent
line and to know how it depends on the black hole spin \citep{svoboda_origin_2012}.
\citet{martocchia_effects_2000} and \citet{miniutti_lack_2003} studied the emissivity profile of an accretion disc illuminated
by a point-like source above the disc and found the profile to be a three-segment broken power-law.
\citet{dovciak_xspec_2014} investigated in detail the dependence of the emissivity profile on the black hole spin and corona height
assuming the lamp-post geometry. Analysis of the emissivity profile of accretion discs irradiated by extended coronae was performed by
\citet{wilkins_understanding_2012} and \citet{gonzalez_probing_2017}.

In most previous studies, the authors made no clear distinction between the illumination profile (the total energy of
illuminating radiation per unit time per unit area) and the emissivity profile of a particular radiative process
(the number of photons emitted per unit time
per unit area). Also, simple assumptions were made for the spectrum of the illuminating radiation. For example, when calculating the
iron K$\alpha$ emissivity profile the spectrum was usually assumed to have
a power-law shape in the energy band most relevant for iron K fluorescent lines.
Unlike the illumination profile, the emissivity profile also depends on the specific radiative process
and is thus sensitive to the local energy spectrum of the illuminating radiation.
In this work, we present the illumination and iron K$\alpha$ emissivity profiles obtained
with a general relativistic (GR) Monte Carlo radiative transfer code \textsc{Monk} \citep{zhang_constraining_2019}.
Thanks to its Monte Carlo nature we are able to calculate not only the flux but
also the spectrum of the incoming illuminating
radiation as measured in the rest frame of the disc fluid for accretion discs illuminated by extended coronae.
As a result, we are able to calculate both the illumination and the 
corresponding emissivity profiles of each radius in the disc.
Recently it has been shown that self-irradiation affects the reflection emission as well \citep[e.g.][]{wilkins_returning_2020,dauser_effect_2022}.
Therefore in this work, we also investigate the illumination and emissivity profiles of self-irradiation. While \citet{wilkins_returning_2020,dauser_effect_2022} studied the returning radiation of the reflection emission due to the coronal emission irradiating the disc, in this manuscript we are investigating the returning radiation of the thermal emission of the disc.

The paper is organized as follows. In Section~\ref{sec:procedure} we illustrate the procedures of the numerical calculations.
In Section~\ref{sec:results} we present the results without taking into account the self-irradiation, 
while in Section~\ref{sec:self-irradiation} we present the results where self-irradiation is taken into consideration.
The results are discussed in Section~\ref{sec:discussion} and summarized in
Section~\ref{sec:summary}.

\section{Procedure}
\label{sec:procedure}
\subsection{Calculating the illumination and emissivity profiles}
We calculate the illumination and emissivity profiles of a razor-thin relativistic Keplerian disc illuminated by an extended corona.
The thin disc extends down to the innermost stable circular orbit (ISCO). We assume that the thin disc follows the Novikov-Thorne
effective temperature profile and there is zero torque at the inner boundary.
We utilise the ``superphoton'' scheme in our simualtion. In this scheme each superphoton
is actually a photon package that consists of many photons with
identical energy and momentum, and is assigned a statistical weight
$w$. The weight $w$ has the physical meaning of the number of photons
generated per unit time as measured in a distant observer’s frame \citep[for details, see][]{zhang_constraining_2019}.
We sample seed superphotons from the thin disc
and propagate the superphotons along null geodesics in the Kerr spacetime.
Each superphoton is characterized by its energy at infinity $E_\infty$, weight $w$, initial position $x_0^\mu$, and 
initial wave vector $k_0^\mu$.
If the superphoton is traveling inside the extended corona, we set the step of raytracing to be much less than the scattering
mean free path. For each step, we evaluate the Compton scattering optical depth. If the
superphoton is scattered, we
sample the energy and momentum of the scattered superphoton assuming the Klein-Nishina differential cross-section.
Finally, we collect superphotons that arrive back to the disc. For each superphoton, we have the following information:
$E_{\infty}$, $w$, and its position $x^\mu$ and wave-vector $k^\mu$ while hitting the disc.

The calculations are performed in the Boyer-Lindquist coordinate system with coordinates $t,r,\theta,\phi$.
We divide the accretion disc into several radial bins, from the ISCO to 100 $\rm GM/c^2$.
The proper area of the $i$-th bin is
\begin{equation}
\begin{split}
\mathcal{S}_i \equiv& \int_0^{2\pi} \int_{r_i}^{r_{i+1}}\gamma\sqrt{g_{rr}g_{\phi\phi}} dr d\phi \\
=&\int_{r_i}^{r_{i+1}} \frac{2\pi\rho \gamma}{\sqrt{\Delta}} \sqrt{r^2 + a^2 + \frac{2a^2 r}{\rho^2}} dr\\
= &\int_{r_i}^{r_{i+1}} 2\pi u^t rdr d\phi,
\end{split}
\end{equation}
where $\rho^2\equiv r^2 + a^2 {\rm cos}^2\theta$, $\Delta \equiv r^2 -2r + a^2$, $\gamma$ is the Lorentzian factor
of the disc fluid as measured by a stationary observer, and $r_i$, $r_{i+1}$ are the lower
and upper boundaries of the $i$-th radial bin, respectively, and
$u^t \equiv dt/d\tau$ is used to translate the time from a distant observer's frame to the local frame:
\begin{equation}
u^t = \frac{1}{\sqrt{1 - 2/r + 4\Omega a / r - \Omega^2 (r^2 + a^2 + 2a^2/r)}},
\end{equation}
where $\Omega$ is the angular velocity of the disc fluid.
In the $i$-th radial bin, the flux of the illuminating radiation in the unit of energy per unit time per unit area 
as measured by an observer co-rotating with the disc fluid is
\begin{equation}
\label{eq:illu_prof}
 \epsilon_i = \frac{\sum w E_{\rm disc} u^t_i}{\mathcal{S}_i},
\end{equation}
where the sum is over all Comptonized photons that strike the thin disc between $r_i$ and $r_{i+1}$,
and $E_{\rm disc}$ is the photon energy measured by
the disc fluid. Denoting the four-velocity of the disc fluid $U^\mu$, we have $E_{\rm disc} = -k^\mu U_\mu E_\infty$.

The probability for a hard X-ray photon to produce an iron K$\alpha$ photon while striking the neutral disc
is proportional to:
\citep[e.g.][]{george_x-ray_1991}:
\begin{equation}
\label{eq:ironk}
P(E) \propto \frac{n_{\rm Fe}\sigma_{\rm Fe}(E)}{\sum n_j\sigma_{\rm j, abs}(E) + n_e\sigma_{\rm sca}(E)},
\end{equation}
if the energy of the hard X-ray photon is above the neutral iron K edge $E_K$ ($\sim 7.12 ~\rm keV$),
where $n_{\rm Fe}$ is the number density of neutral iron, 
$n_j$ is the number density of the $j$-th ion species, $n_e$ is the number density
of electrons, $\sigma_{Fe}$ is the photon-ionization cross section of the neutral iron,
$\sigma_{j,abs}$ is the photon-ionization cross section of the $j$-th ion species,
and $\sigma_{\rm sca}$ is the Compton scattering cross section.
As $\sigma_{Fe}\propto E^{-3}$ while $\sigma_{\rm sca}$
is varying slowly with energy, $P(E)$ decreases with energy
rapidly and the photons with energy just above the iron K edge are most essential for the production of 
iron K$\alpha$ photons \citep{george_x-ray_1991}.

In the $i$-th radial bin, the emissivity of iron K$\alpha$ photon, 
i.e. the number of iron K$\alpha$ photons produced per unit area per unit time
in the local frame is 
\begin{equation}
\label{eq:emis_prof}
 \varepsilon_i = \frac{\sum wP(E_{\rm disc})u^t_i}{\mathcal{S}_i},
\end{equation}
where the sum is over all superphotons with $r_i \leq r < r_{i + 1}$ and $E_{\rm disc} \geq E_K$.

In this paper, we assume that the elements in the disc atmosphere are neutral and have solar abundance \citep{grevesse_standard_1998}.
The photon ionization cross sections are calculated with analytical formulae by \citet{verner_analytic_1995}. 
For Compton scattering we assume the Klein-Nishina scattering cross section.


\subsection{Relation between the illumination and emissivity profiles}
\label{sec:relation}
Let us assume that the spectra of the illuminating radiation at different radii on the disc differ only by different
normalizations and redshift (which is the case for an isotropic, lamp-post corona).
In this case, we can write the photon spectrum of the illuminating radiation as observed by the disc fluid located at radius $r$ as
\begin{equation}
 N(E, r) = F(r) f\left(\frac{E}{g(r)}\right)\frac{1}{g(r)},
\end{equation}
where $F(r)$ is the radial distribution of the illuminating photons and is a function of $r$ only, $f(E)$ is the energy spectrum
of the illuminating radiation in the rest frame of the corona, and
$g(r)$ is the redshift factor $g\equiv E_{\rm disc}/E_{\rm corona}$. Then the illuminating radiation profile is
\begin{equation}
 \epsilon(r) = \int_0^{+\infty} E N(E, r) dE \propto F(r)g(r),
\end{equation}
regardless of the spectral shape. The iron K$\alpha$ emissivity profile
\begin{equation}
 \varepsilon(r) = \int_{E_K}^\infty  N(E, r)P(E) dE = \frac{F(r)}{g(r)} \int_{E_K}^\infty P(E) f\left(\frac{E}{g(r)}\right) dE.
\end{equation}
For Comptonized spectrum the high energy spectrum can usually be described by a cut-off power-law function
$f(E)=E^{-\Gamma}e^{-E/E_{\rm cut}}$ for $E\geq E_0$, where $\Gamma$ is the photon index, $E_{\rm cut}$ is the high-energy cut-off 
energy, and $E_0$ is the energy of the low-energy cut-off. If $E_0\leq E_K / max(g(r))$, then
\begin{equation}
 \varepsilon(r) = F(r)g^{\Gamma-1}(r) \int_{E_K}^\infty P(E) {E}^{-\Gamma} e^{-E/g(r)E_{\rm cut}} dE.
\end{equation}
For the iron K$\alpha$ emissivity profile, as $P(E)$ decrease rapidly with energy, the integral is not sensitive to $g(r)$, therefore
\begin{equation}
 \varepsilon(r) \propto F(r)g^{\Gamma-1}(r),
\end{equation}
and 
\begin{equation}
\label{eq:eq11}
 \frac{\varepsilon(r)}{\epsilon(r)} \propto g^{\Gamma-2}(r).
\end{equation}
We can see that whether the emissivity profile is steeper than the illumination profile
depends on the value of $\Gamma$ and the shape of $g(r)$.

\section{Results}
\label{sec:results}
\subsection{Stationary spherical corona in AGNs}
\label{sec:sphcorona}
In this section, we present the illumination and emissivity profiles of accretion discs illuminated by a spherical corona above the disc. In the reminder of this paragraph we list the fiducial values of the simulations.
The rapidly rotating ($a=0.998$) black hole with a mass of $10^7~\rm M_\odot$ is accreting with a mass accretion rate of
$4.32\times10^{23}~\rm g~s^{-1}$. Assuming the radiative efficiency to be $0.354$ for a spin $0.998$ black hole, the bolometric
luminosity is expected to be $\sim10\%$ the Eddington luminosity.
The spherical corona is stationary, i.e. the fluid has the same angular velocity as a zero angular momentum observer.
The temperature of the corona is $100~\rm keV$, and the Thomson optical depth of the corona $\tau_{\rm T} \equiv \sigma_{T} n_e R_c = 0.2$,
where $\sigma_{\rm T}$ is the Thomson scattering cross section, and $R_c$ is the radius of the corona. Given $\sigma_{\rm T}$, we immediately obtain the electron density of the corona. For each tiny step of raytracing, the procedure for evaluating the scattering optical depth can be found in Sec. 2.6.1 of \citet{zhang_constraining_2019}.

\subsubsection{Dependence on the corona size}
\begin{figure}
 \includegraphics[width=\columnwidth]{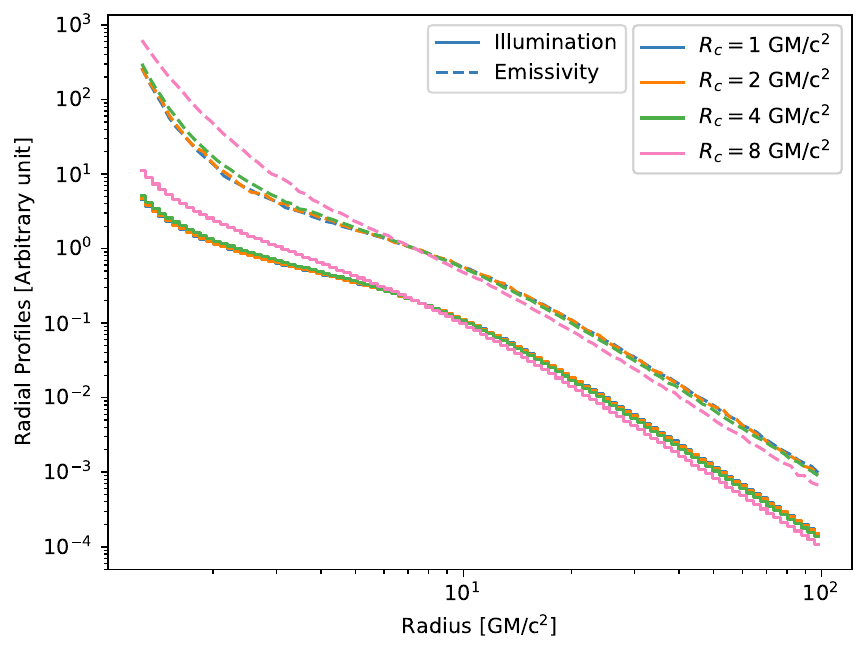}
 \caption{The illumination (solid lines) and emissivity (dashed lines) profiles of
 accretion discs illuminated by stationary spherical coronae of different sizes. Each corona is located on the symmetry axis of a $10^7~\rm M_\odot$ black hole,
 $10~\rm GM/c^2$ above the equatorial plane.\label{fig:sphere_size}}
\end{figure}

In Fig.~\ref{fig:sphere_size}, we present the illumination and emissivity profiles for spherical coronae of different sizes.
The center of the corona is located on the black hole rotation axis, $10~\rm GM/c^2$ above the equatorial plane.
For all corona sizes the emissivity profile seems to be quite steep in the innermost region of the accretion disc (below
$\sim2-3~\rm GM/c^2$), and becomes shallower as the radius increases. Then beyond $\sim 10~\rm GM/c^2$, it steepens again.
We fit the emissivity profile in different regions on the accretion disc with a power-law model (i.e. $\varepsilon(r) \propto r^{-q}$) using
the least squared method, and summarize the results in Table~\ref{tab:sphere_size}. In the innermost region of the accretion disc
($r\leq 2~\rm GM/c^2$), the emissivity profile becomes steeper as the corona radius decreases. 
Far away from the black hole ($r\geq 20 ~\rm GM/c^2$), the indices are roughly consistent with the Newtonian value of $3$.
In the Newtonian case, for a lamp-post corona we expect illumination profile to have the following radial profile:
\begin{equation}
 \epsilon(r) \propto \frac{1}{(h^2 + r^2)^{3/2}},
\end{equation}
where $h$ is the height of the corona, which has a power-law shape with an index of 3 at large radii and flattens in the inner region. This explains the radial profile
in the intermediate and outer regions where the gravitational field is not profound.
In the innermost region, the light-bending effect of the strong gravitational field leads to a steep profile.

\begin{table}
\begin{center}
\caption{The fitted indices of the illumination and emissivity profiles,
for accretion discs illuminated by spherical and slab coronae of various radii.\label{tab:sphere_size}}
\begin{tabular}{|lccccc|}
\hline
Geometry & $R_c$ & \multicolumn2c{Illumination} & \multicolumn2c{Emissivity} \\
& $[{\rm GM/c^2}]$ & $r < 2$ & $r> 20$ & $r<2$ & $r>20$ \\
\hline
\multirow{4}{*}{Spherical} & 1 & 2.92 & 3.06 & 6.55 & 2.97 \\
&     2 & 2.92 & 3.06 & 6.49 & 3.00 \\
 &   4 & 2.96 & 3.07 & 6.32 & 2.98 \\
 &   8 & 3.27 & 3.09 & 5.67 & 3.02 \\
\hline
\hline
\multirow{4}{*}{Slab} & 1 & 2.84 & 3.06 & 6.09 & 2.95 \\
 & 2 & 2.84 & 3.05 & 4.82 & 2.97 \\
 & 4 & 2.92 & 3.08 & 4.42 & 3.00 \\
 & 8 & 3.00 & 3.31 & 4.09 & 3.26 \\
\hline

\end{tabular}
\end{center}
\end{table}

The illumination profile has a similar three-segment broken power-law shape
but in the innermost region it is shallower than the emissivity profile.
We also fit the profiles and present the results in Table~\ref{tab:sphere_size}.
We measure the photon index of the illuminating radiation in the energy band of 20--100 keV and find it to be $\sim3$.
According to \citet{dovciak_xspec_2014}, 
in the innermost region of the accretion disc $g(r)$ increases with decreasing radius.
Therefore, the result that the emissivity profile is steeper than the illumination profile
is consistent with Eq.~\ref{eq:eq11}.

\subsubsection{Dependence on the corona height}
\begin{figure}
 \includegraphics[width=\columnwidth]{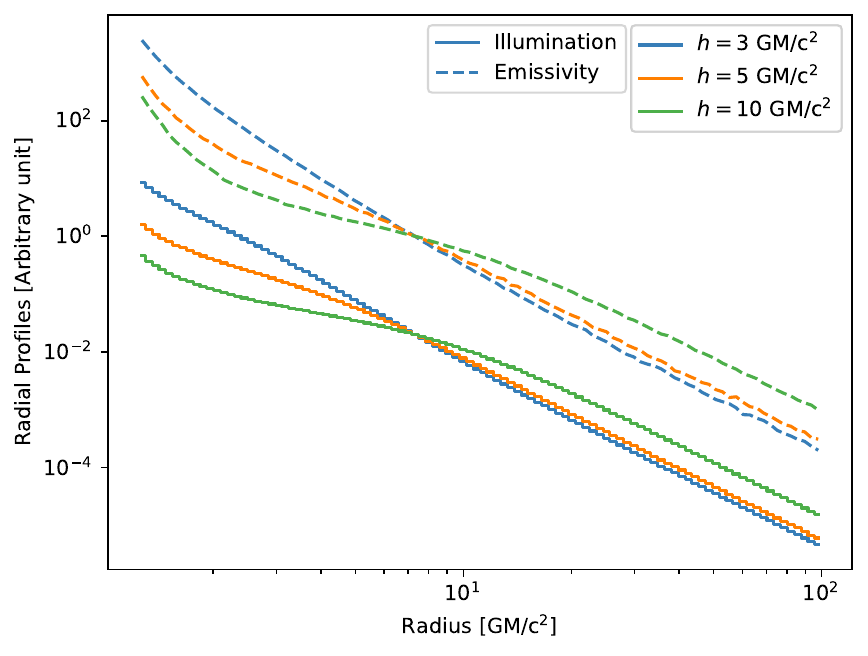}
 \caption{The same with Fig.~\ref{fig:sphere_size}, but for coronae with radius of $1~\rm GM/c^2$ and various heights
 above the equatorial plane.
\label{fig:sphere_height}}
\end{figure}

In Fig.~\ref{fig:sphere_height}, we present the illumination and emissivity profiles for spherical coronae
located at different heights above the equatorial plane. The radius of the coronae is 1 $\rm GM/c^2$ for all cases.
We fit the emissivity and illumination profiles in different regions and summarize the results in
Table~\ref{tab:sphere_height}.
In the innermost region, as the height increases, the emissivity profile becomes steeper.
The illumination profile steepens towards lower height, which is similar with lamp-post corona \citep{dovciak_xspec_2014}, and
is opposite to the emissivity profile.

\begin{table}
\begin{center}
    \caption{The fitted indices of the illumination and emissivity profiles,
    for accretion discs illuminated by spherical and slab coronae of various heights in the lamp-post geometry.\label{tab:sphere_height}}
\begin{tabular}{|lccccc|}
    \hline
Geometry & $h$ & \multicolumn2c{Illumination} & \multicolumn2c{Emissivity} \\
& $[{\rm GM/c^2}]$ & $r < 2$ & $r> 20$ & $r<2$ & $r>20$ \\
\hline
\multirow{3}{*}{Spherical} & 3 & 3.52 & 3.13 & 5.80 & 3.15 \\
& 5 & 3.01 & 3.12 & 5.90 & 3.13 \\
& 10 & 2.92 & 3.06 & 6.55 & 2.97 \\
\hline
\hline

 \multirow{3}{*}{Slab, $R_c=1~\rm GM/c^2$} & 3 & 2.30 & 3.19 & 4.72 & 3.28 \\
& 5 & 2.78 & 3.17 & 5.60 & 3.11 \\
& 10 & 2.84 & 3.06 & 6.09 & 2.95 \\
    \hline
\multirow{3}{*}{Slab, $R_c=8~\rm GM/c^2$} & 3 & 2.77 & 2.98 & 3.19 & 2.97 \\
&  5 & 3.15 & 3.02 & 3.74 & 3.02 \\
&  10 & 3.00 & 3.31 & 4.09 & 3.26 \\
  \hline
   \hline
\end{tabular}
\end{center}
\end{table}

\subsubsection{Dependence on other parameters}
\begin{figure}
 \includegraphics[width=\columnwidth]{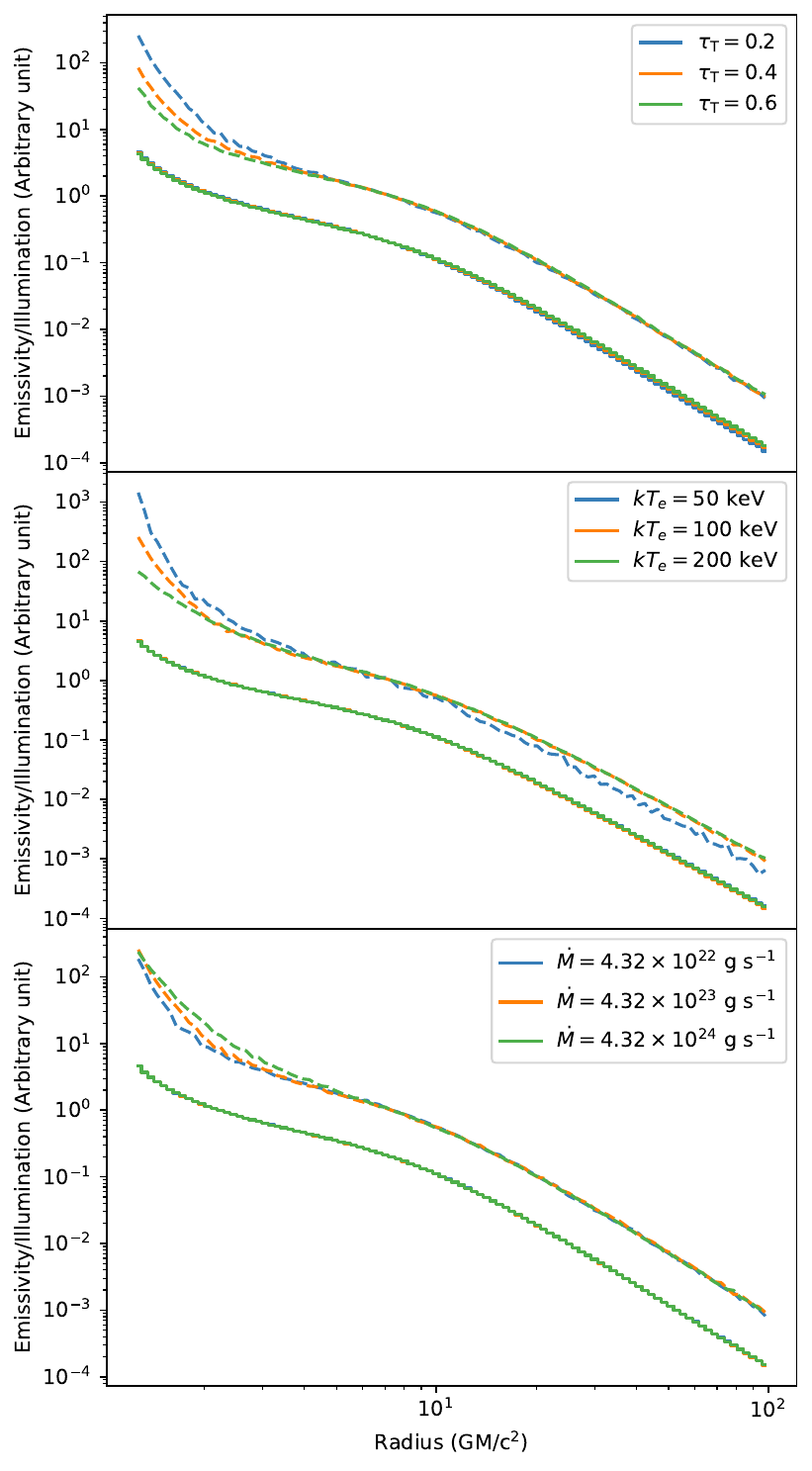}
 \caption{The illumination (solid) and emissivity (dashed) profiles of coronae with varying non-geometrical parameters. Top panel: the profiles of different optical depth of the corona. Middle: the profiles assuming various coronal temperature. Bottom: the profiles assuming various mass accretion rates of the disc.\label{fig:sphere_spectral}}
\end{figure}

We also investigate the effect of other, non-geometrical parameters of the corona on the illumination and emissivity profiles. We assume corona at a height of $10~\rm GM/c^2$ and radius of $1~\rm GM/c^2$. We start with varying the optical depth of the corona while fixing other parameters at the fiducial values. In the top panel of Fig.~\ref{fig:sphere_spectral}, we present the profiles for various optical depthes of the corona. The illumination profiles are identical for different optical depth, which is expected as the illumination profile is almost solely determined by the coronal geometry. In the innermost region, the emissivity profiles get shallower as the optical depth increases, consistent with our calculation in Sec.~\ref{sec:relation} (especially Eq.~\ref{eq:eq11}), as the photon index decreases with the optical depth. Similar analysis is done for the coronal temperature and the mass accretion rate, the results of which are shown in the middle and bottom panels of Fig.~\ref{fig:sphere_spectral}, respectively. Again, the illumination profile is independent of these two parameters. The emissivity profile gets shallower with the increasing coronal temperature, due to the same reason with the dependence on the optical depth, that the spectrum hardens as the coronal temperature increases. The emissivity profile is also weakly dependent on the mass accretion rate, probably due to the mass accretion rate affecting the low-energy cut-off of the primary emission.

\subsection{Co-rotating slab corona in AGNs}
In this section, we present the results for accretion discs irradiated by
slab coronae above the disc. The slab coronae are
co-rotating with the underlying accretion disc. The thickness of the slab is $2~\rm GM/c^2$. The optical depth of the
corona along the vertical direction is $\tau_{T} \equiv n_e \sigma_{T} h_c/2 = 0.2$, where $h_c$ is the thickness of the corona,
and the temperature of the corona is
$100~\rm keV$. The black hole and thin disc have the same properties as in the previous subsection.

\subsubsection{Dependence on the corona size}
\begin{figure}
 \includegraphics[width=\columnwidth]{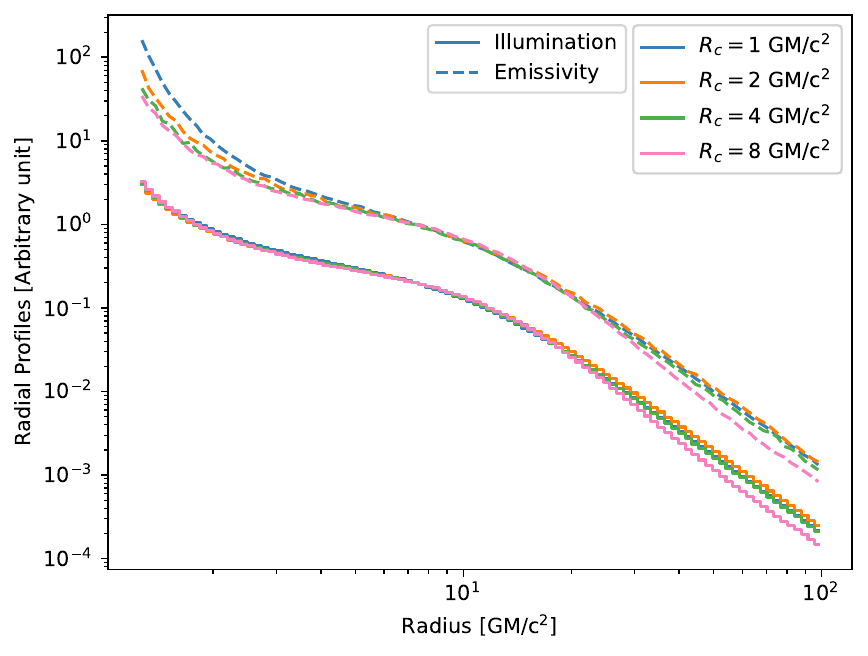}
 \caption{The illumination and emissivity profiles of accretion discs illuminated by
 co-rotating slab coronae of different sizes. The coronae have a thickness of $2~\rm GM/c^2$ and are located on the symmetry axis of a $10^7 ~\rm M_\odot$ black hole, 
 $10~\rm GM/c^2$ above the equatorial plane. The various radial extents of the slab coronae along the horizontal direction are indicated in the figure.\label{fig:slab_size}}
\end{figure}

In Fig.~\ref{fig:slab_size}, we present the illumination and emissivity profiles for accretion discs irradiated by
slab coronae of different sizes. 
Compared with the spherical coronae, the emissivity profile is more sensitive to the corona size, 
with shallower emissivity from more extended corona. 
Similarly with spherical coronae, the illumination profile is shallower than the emissivity profile given
the same corona size. We fit the profiles and tabulate the results in Table~\ref{tab:sphere_size}.



\subsubsection{Dependence on the corona height}
\begin{figure}
 \includegraphics[width=\columnwidth]{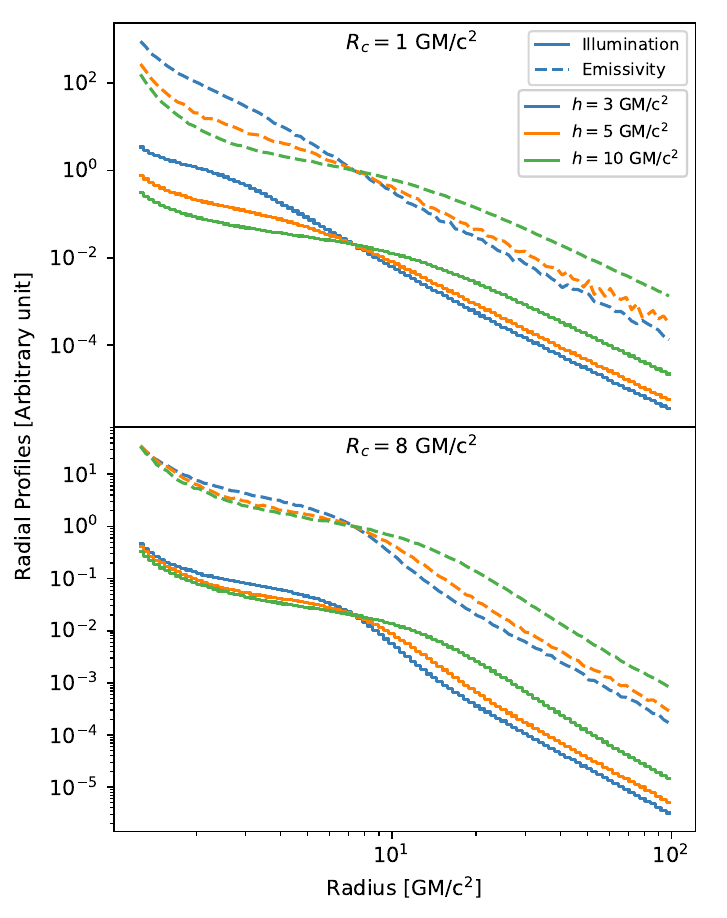}
 \caption{The same with Fig.~\ref{fig:slab_size}, but for coronae with the same radial extent but various heights
 above the accretion discs. In the upper and lower panels we present the profiles for corona radii of $1$ and $8~\rm GM/c^2$, respectively.
 \label{fig:slab_height}}
\end{figure}

In Fig.~\ref{fig:slab_height}, we present the illumination and emissivity profiles for accretion discs irradiated by
slab coronae located at different heights above the accretion disc. As the height increases, both the emissivity and the illumination
profiles become steeper in the innermost region. The fitting results are tabulated in Table~\ref{tab:sphere_height}.

\subsection{Stationary spherical coronae in BHXRBs}
\begin{figure}
 \includegraphics[width=\columnwidth]{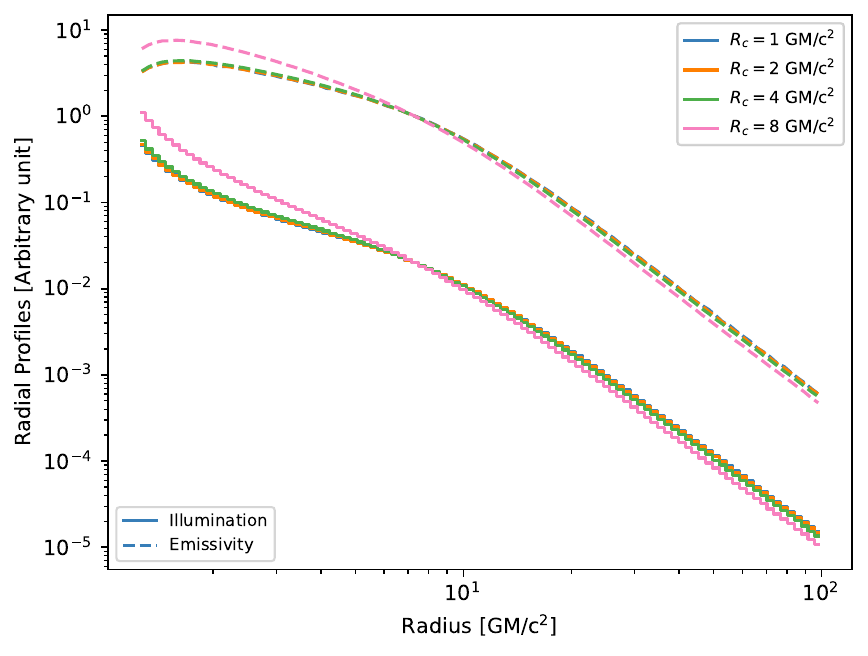}
 \caption{The same with Fig.~\ref{fig:sphere_size}, but for a $10~\rm M_\odot$ black hole.
 \label{fig:sphere_size_xrb}}
\end{figure}

In Fig.~\ref{fig:sphere_size_xrb}, we present the illumination and emissivity profiles of accretion discs in
BHXRBs illuminated by spherical coronae. The parameters are the same as for AGNs in Section~\ref{sec:sphcorona}, but for a $M=10~\rm M_\odot$ black
hole accreting at a rate of $4.32\times10^{17}~\rm g~s^{-1}$ (corresponding to 10\% Eddington luminosity).
The emissivity profile is distinct compared
with the emissivity profile of AGNs (Fig.~\ref{fig:sphere_size}) in that the profile is shallower than the Newtonian case below
$\sim10~\rm GM/c^2$, and the emissivity even decreases towards lower radius below $\sim1.6~\rm GM/c^2$.

\begin{figure}
 \includegraphics[width=\columnwidth]{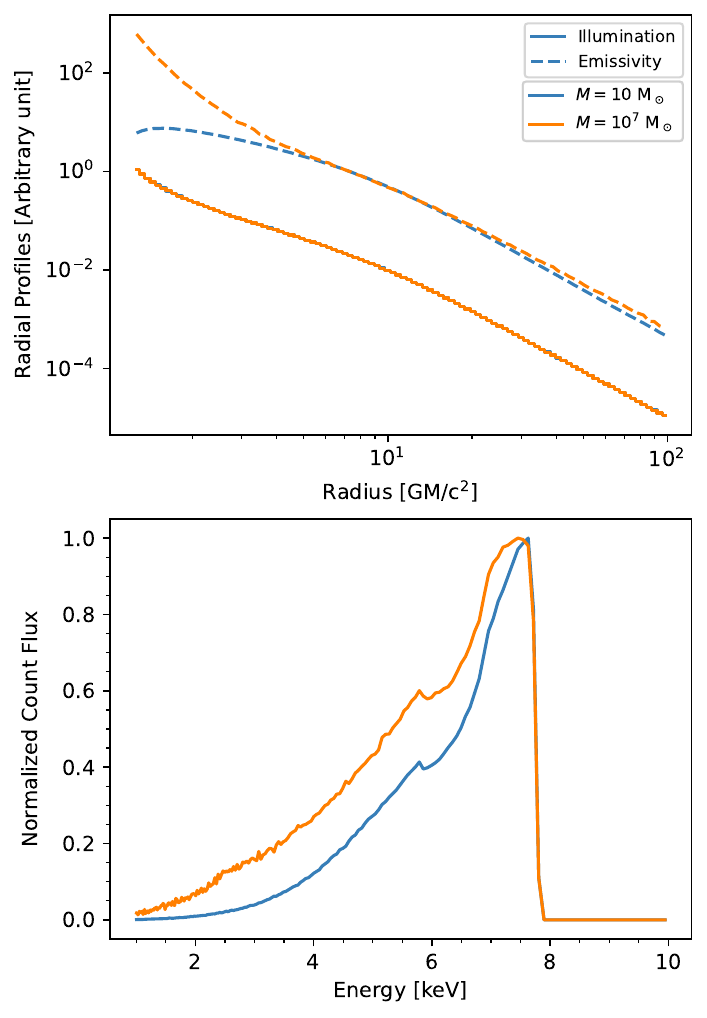}
 \caption{The upper panel: a comparison between the emissivity (dashed) and illumination profiles of accretion discs in a BHXRB (blue)
 and in an AGN (orange). In both cases the accretion discs are illuminated by a stationary spherical corona with a height of
 $10~\rm GM/c^2$ and a radius of $1~\rm GM/c^2$. The lower panel: the line profiles corresponding to the emissivity profiles in the upper panel, 
 for an observer at an inclination of $60^\circ$.
\label{fig:agnvsxrb}}
\end{figure}

To highlight the distinction in the emissivity profile, in the upper panel of Fig.~\ref{fig:agnvsxrb} we compare
the profiles of accretion discs around $10~\rm M_\odot$ and $10^7~\rm M_\odot$ black holes. The illumination profiles are almost
identical while the difference in the iron K$\alpha$ emissivity profile is obvious.
An immediate consequence is that the iron line profile becomes different.
We calculate the iron line profile seen by an observer at an inclination of $60^\circ$,
assuming isotropic iron line emission in the rest frame of the disc. The results are presented in the lower
panel of Fig.~\ref{fig:agnvsxrb}. Assuming the same corona geometry, the iron line profiles from AGNs and XRB are clearly different, with
the latter weaker in the red wing.

\begin{figure}
 \includegraphics[width=\columnwidth]{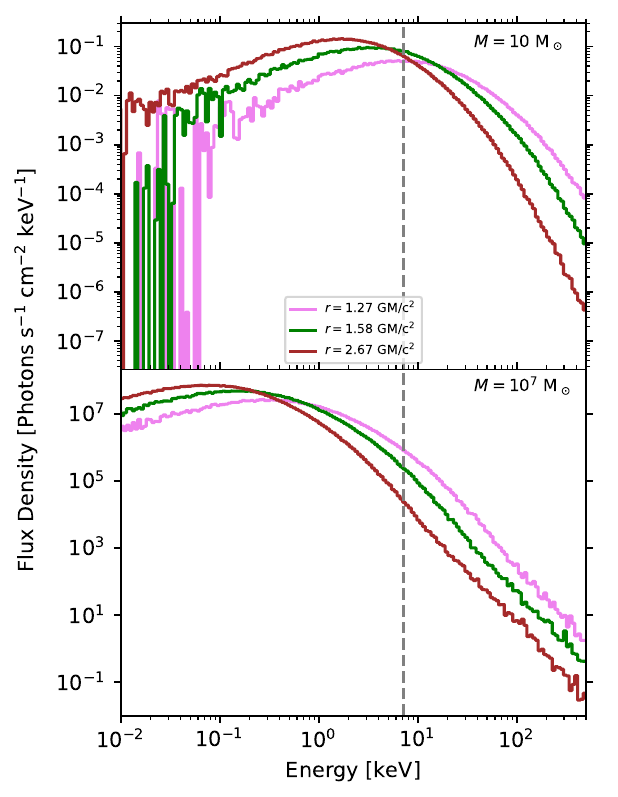}
 \caption{The energy spectra of the illuminating radiation observed by disc fluid located at different radii. The spherical corona is
 located $10~\rm GM/c^2$ above the accretion disc and has radius of $1~\rm GM/c^2$. The spectra at different radii are plotted
 in different colors.
 The vertical dashed line
 indicates the location of neutral iron K edge at 7.12 keV. In the upper and lower
 panels we present the spectra for accretion discs around black holes with masses of $10$ and $10^7~\rm M_\odot$,
 respectively.\label{fig:enspec}}
\end{figure}

To find out the reason that causes the difference, we calculate the spectra of the
illuminating radiation as observed by the disc fluid at radii of $1.27$, $1.58$, and $2.67~\rm GM/c^2$, respectively, and present the
results in Fig.~\ref{fig:enspec}.
The spectra can be described as a broken power-law with a low-energy spectral break roughly at the temperature of the seed photon.
While the break energy should stay the same in the rest frame of the corona, the change of the redshift factor across the accretion disc results in different break energies of the illuminating radiation.
For $M=10^7~\rm M_\odot$, the break energy is always lower than $E_K$.
In this case, a higher redshift simply leads to more X-ray photons above $E_K$;
whereas for the $10~\rm M_\odot$ case, the break energy of the photon spectra is in the range of $2-10~\rm keV$,
resulting in a distinct emissivity profile.
This indicates that the energy spectra of the hard X-ray radiation could substantially affect the emissivity profile.

\begin{figure}
 \includegraphics[width=\columnwidth]{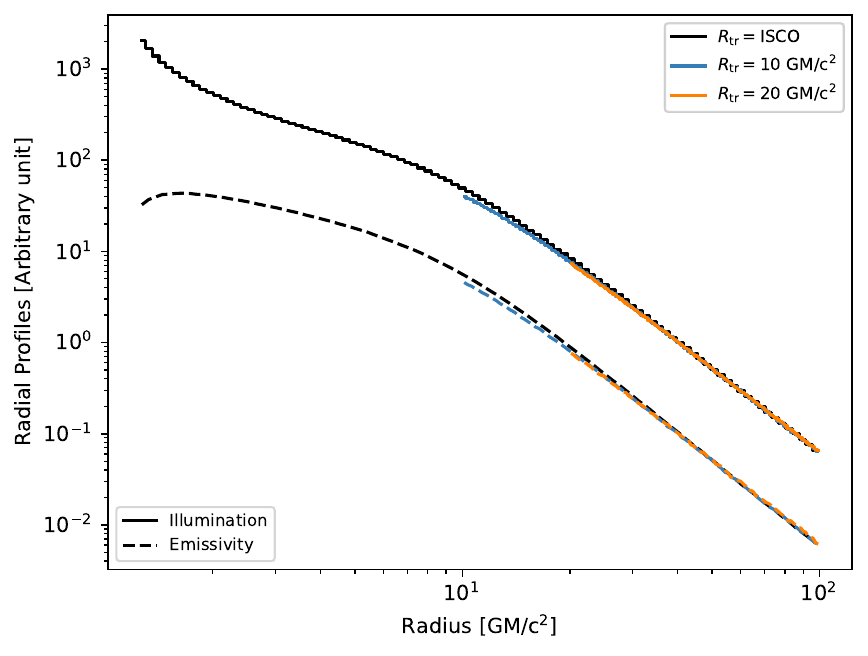}
 \caption{The same with Fig.~\ref{fig:sphere_size_xrb}, but for various truncation radii $R_{\rm tr}$, as indicated in the plot.\label{fig:sphere_xrb_rtr}}
\end{figure}

In the most popular scenario of the spectral states of BHXRBs, the standard disc during the hard spectral state is truncated at a radius larger than the ISCO \citep[e.g.][]{esin_advection-dominated_1997}. We therefore perform simulations to assess the illumination and emissivity profiles for truncated discs. In Fig.~\ref{fig:sphere_xrb_rtr} we present the profiles for discs truncated at radii of $10$ (blue) and $20~\rm GM/c^2$ (orange), and compare the results with discs truncated at the ISCO (black). Beyond the truncation radius, the illumination and emissivity profiles of truncated disc seem to be almost the same with the profiles of accretion discs truncated at ISCO, indicating that the profiles are rather insensitive to the truncation radius.

\section{The effect of self-irradiation}
\label{sec:self-irradiation}
So far we have considered the irradiation of the corona emission only. In fact, due to the light-bending effect of the
strong gravitational field in the vicinity the black hole,
the accretion disc, especially the inner part of it, is also irradiated by photons from the opposite side of the disc, which is commonly referred to as self-irradiation.
In this section we investigate the illumination and emissivity profile after taking the self-irradiation into account.
The calculations are done utilizing equations~\ref{eq:illu_prof} and \ref{eq:emis_prof}, in a similar fashion with the corona
radiation, except that the sum is now over all the self-irradiating photons.

\subsection{BHXRB}
\begin{figure}
 \includegraphics[width=\columnwidth]{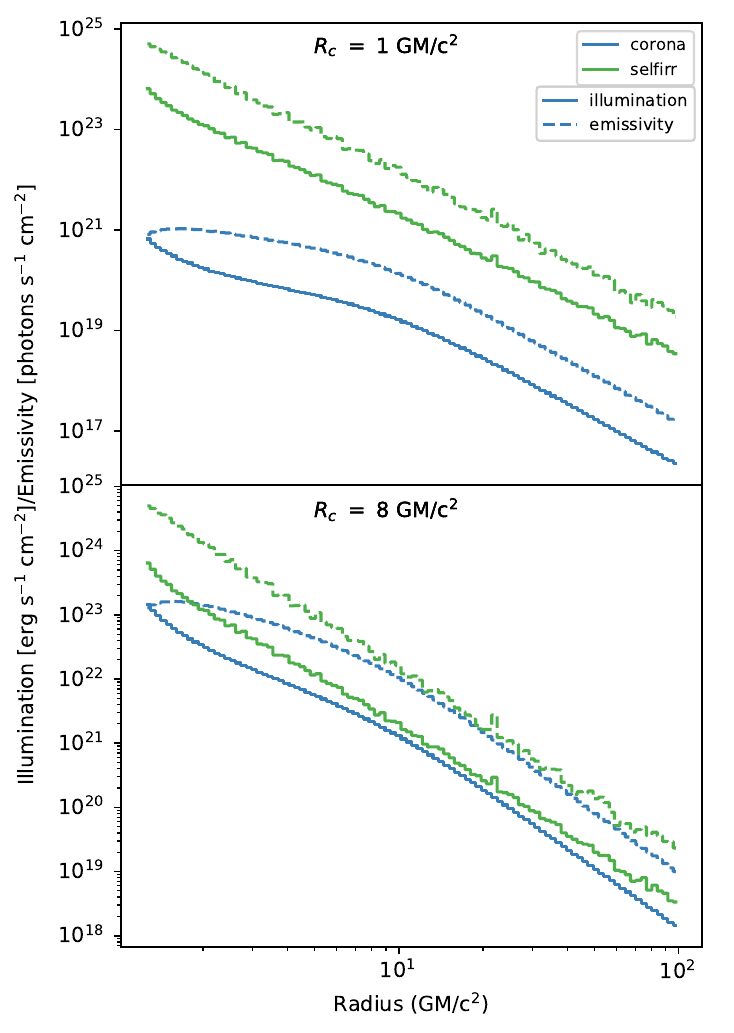}
 \caption{The illumination and iron K$\alpha$ emissivity profiles of the corona 
 emission in blue and of the self-irradiation in cyan, for a BHXRB with a black hole mass of $10~\rm M_\odot$.
 Note that here the profiles are in physical units.
 The profiles for coronae with radii of $1$ and $8~\rm GM/c^2$ are presented in the upper and lower panels,
 respectively.\label{fig:selfirr_xrb}}
\end{figure}

In Fig.~\ref{fig:selfirr_xrb}, we present the illumination and emissivity profiles of the self-irradiation in a BHXRB and compare them with the profiles of the corona emission. For a compact corona with radius of $1~\rm GM/c^2$ (the upper panel), the illumination is dominated by self-irradiation, while the corona emission is weaker by $\sim2$ orders of magnitudes. In contrast with the three-segment-power-law shape of the illumination profile of the corona emission, the illumination profile of the self-irradiation seems to have a power-law shape without obvious breaks. The iron K$\alpha$ emissivity profile also has a power-law shape down to the innermost region of the accretion disc. The profiles for a more extended corona ($R_c=8~\rm GM/c^2$) are plotted in the lower panel of Fig.~\ref{fig:selfirr_xrb}. The illumination from the corona and the self-irradiation are comparable beyond $\sim 10~\rm GM/c^2$, whilst in the inner region the self-irradiation is more important. We would also expect the relative importance of the self-irradiation to be smaller if the disc is truncated, as the self-irradiation is due to the strong gravitational field close to the black hole.

\begin{figure}
 \includegraphics[width=\columnwidth]{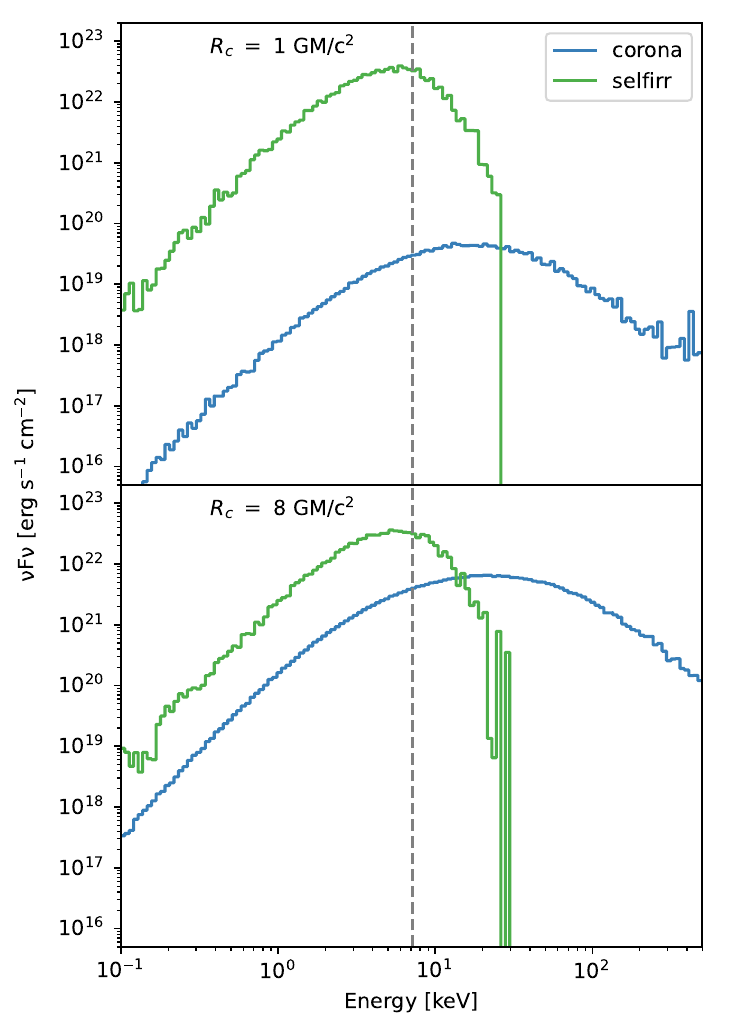}
 \caption{The energy spectra of the corona emission (blue) and the self-irradiation (cyan),
as seen by the disc fluid located at a radius of $2.67~\rm GM/c^2$, for a BHXRB with a black hole mass of $10~\rm M_\odot$. The spectra for corona radii of $1$ and $8~\rm GM/c^2$ are plotted in the
upper and lower panels, respectively. In both panels, the energy of the iron K edge at $7.12 ~\rm keV$ is indicated by a vertical dashed line.\label{fig:selfirr_enspec_xrb}}
\end{figure}

\begin{figure}
 \includegraphics[width=\columnwidth]{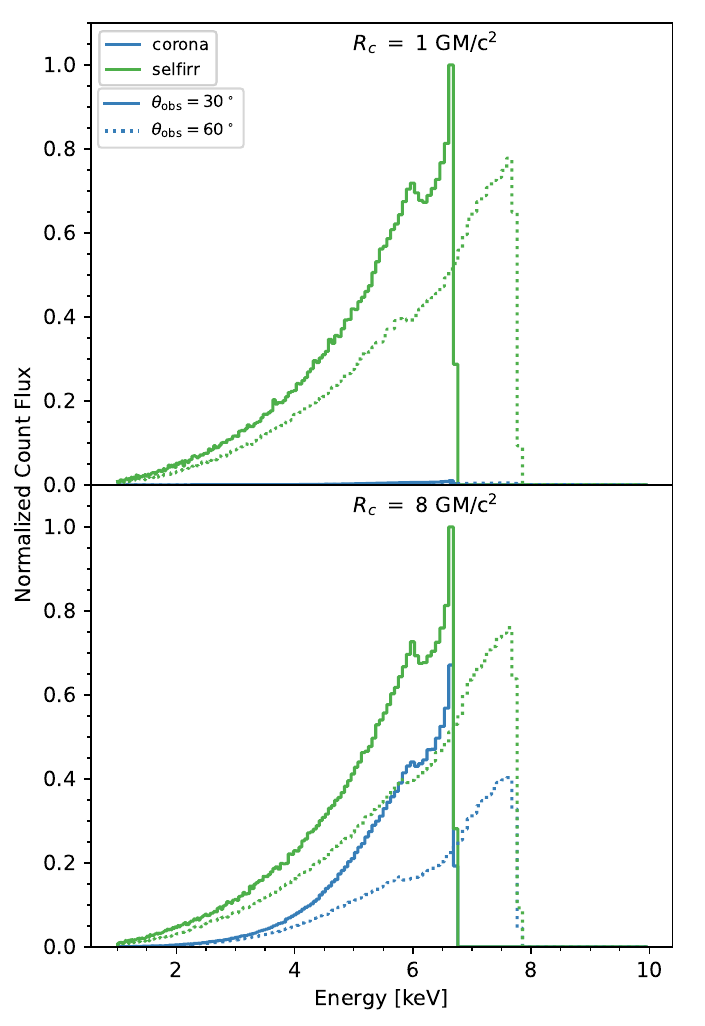}
 \caption{The profiles of the iron line as observed by observers at inclinations of $30^\circ$ (solid lines) and $60^\circ$
 (dashed lines), corresponding to the emissivity profiles in Fig.~\ref{fig:selfirr_xrb}.
 The contributions of the corona and self-irradiation are plotted in blue and cyan colors, respectively.
 The results for corona radii of $1$ and $8~\rm GM/c^2$ are plotted in the upper and lower panels, respectively.\label{fig:selfirr_xrb_iron}}
\end{figure}

In Fig.~\ref{fig:selfirr_enspec_xrb}, we present the energy spectra of the corona emission and self-irradiation as seen by disc fluid located at a radius of $2.67~\rm GM/c^2$. For corona radius of $1~\rm GM/c^2$ (the upper panel), the self-irradiation is brighter than the corona emission below $\sim 30~\rm keV$. As the corona becomes more extended with a radius of $8~\rm GM/c^2$ (the lower panel), the self-irradiation barely changes but the corona emission becomes much stronger, therefore the critical energy below which the self-irradiation dominates drops to $\sim15~\rm keV$, lower than the case of $R_c=1~\rm GM/c^2$, but still above the iron K edge. Therefore, we expect that in BHXRBs the self-irradiation plays an important role in the iron K emission process. This is demonstrated in Fig.~\ref{fig:selfirr_xrb_iron}, in which the energy spectra of the iron K line are presented.

\subsection{AGNs}
\begin{figure}
 \includegraphics[width=\columnwidth]{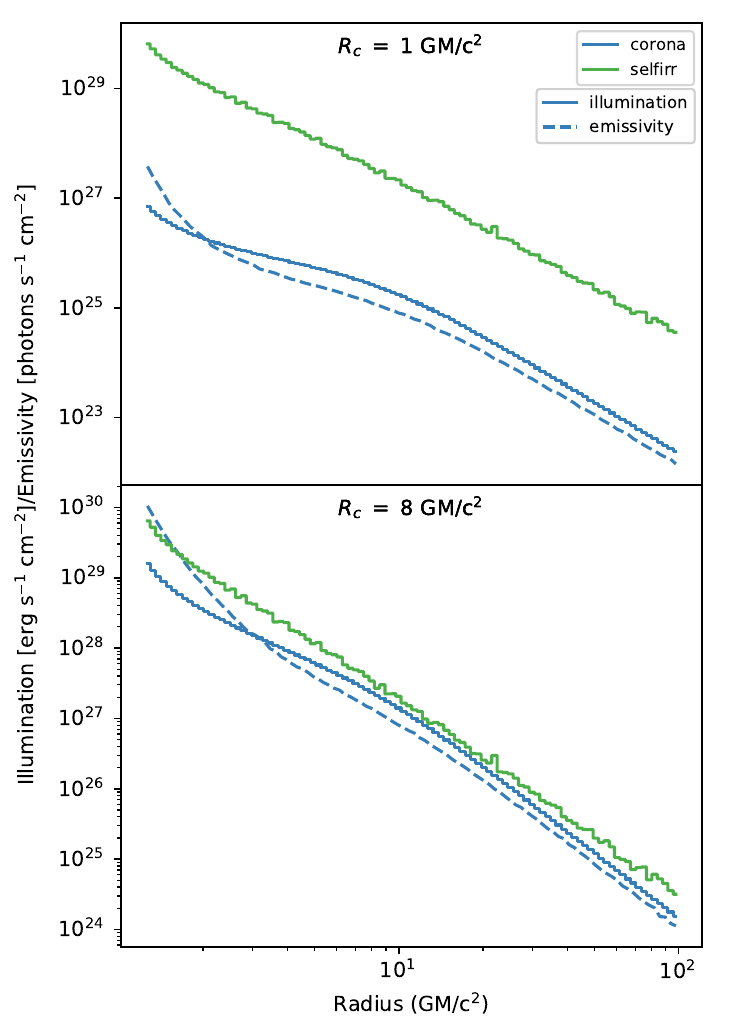}
 \caption{The same with Fig.~\ref{fig:selfirr_xrb},
 but for an AGN with black hole mass of $10^7~\rm M_\odot$. The self-irradiation is barely contributing to the iron line emissivity.
\label{fig:selfirr}}
\end{figure}

\begin{figure}
 \includegraphics[width=\columnwidth]{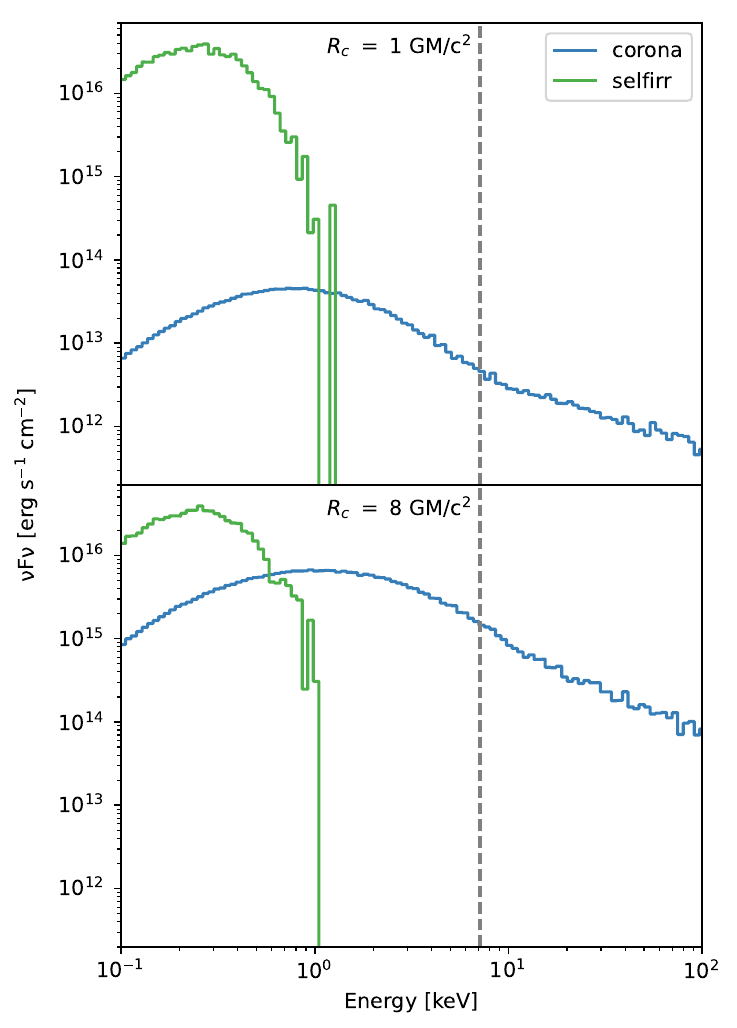}
 \caption{The same with Fig.~\ref{fig:selfirr_enspec_xrb},
 but for an AGN with a black hole mass of $10^7~\rm M_\odot$.\label{fig:selfirr_enspec}}
\end{figure}

In Fig.~\ref{fig:selfirr}, we plot the spectra of self-irradiation in AGNs. The behavior of the illumination profile is the same with
the BHXRB case. In contrast, the self-irradiation is barely contributing to the production of the iron K line,
and the iron K emissivity is dominated by the corona emission.
The reason is that the cooler self-irradiation in AGNs cannot reach the iron K edge.
This can be clearly seen in Fig.~\ref{fig:selfirr_enspec},
where we present the energy spectra of the corona emission as well as self-irradiation. The $\nu F\nu$ spectrum of
self-irradiation is peaked at $\sim0.2 ~\rm keV$ and its contribution disappear above $\sim 1~\rm keV$.
In contrast with the iron K$\alpha$ emission,
the soft excess component in the reflection spectrum would be affected by the self-irradiation, as discussed in more details in Section~\ref{sec:soft_excess}.

\section{Discussion}
\label{sec:discussion}
\subsection{A summary of our results}
The trajectory of a photon in vacuum is independent of the energy of the photon. As a result, given two disc-corona systems with the same black hole spin and geometries (in geometrical units) but with different mass, we would expect nearly the same radial distribution of the Comptonized photons that strike the underlying disc, independent of the mass of the black hole.

However, the reflection emission depends not only on the flux but also on the spectral shape of the illuminating coronal emission. One example is the iron K$\alpha$ line in the reflection spectrum (see equation~\ref{eq:ironk}). The energy spectrum of the corona radiation as seen by the disc depends on the temperature of the disc, which is sensitive to the mass of the black hole and the mass accretion rate. This is demonstrated in Section~\ref{sec:results}, where we show that the discs in AGNs and XRBs have identical illumination profiles but distinct iron $K\alpha$ emissivity profiles.

We note that we make simple assumptions on the corona and the disc. For instance, we assume a razor-thin disc, while in reality the disc should have finite thickness \citep[e.g.][]{taylor_exploring_2018}. We also assume spherical or slab coronae, in contrast to vertically extended corona that are studied in a few works \citep[e.g.][]{chainakun_investigating_2017,lucchini_investigating_2023}. A more formal treatment of the reflection spectrum should also take into account the physical properties, e.g. the density of the disc \citep[e.g.][]{jiang_xmm-newton_2022}. While in future works we are going to investigate disc and corona with more realistic assumptions, we emphasize that the difference in the illumination and emissivity profiles we find in this work is more related with the fact that we consider the energy spectrum of the primary emission as received by the disc, and therefore we would tentatively expect that this effect is less sensitive to the geometrical parameters of the disc or the corona and the density of the disc.



\subsection{Self-irradiation and the soft excess}
\label{sec:soft_excess}
Below $\sim 1-2~\rm keV$, the soft X-ray spectra of many AGNs exhibit excess with respect to the power-law continuum,
known as ``soft excess''. This feature was first noticed by \citet{arnaud_exosat_1985} and \citet{singh_observations_1985},
and later found in many
Seyfert galaxies \citep[e.g.][]{pounds_soft_1986,turner_variability_1988,walter_ultraviolet_1993,page_xmm-newton_2004,
porquet_xmm-newton_2004,crummy_explanation_2006,bianchi_caixa:_2009-1,miniutti_xmm_2009}.

Currently there are two most popular explanations for the soft excess in AGNs: warm corona and ionized reflection.
In the warm corona scenario, the soft excess is due to a ``warm'' corona Compton up-scattering optical/UV photons 
originating from the underlying thin disc
\citep[e.g.][]{czerny_constraints_1987,magdziarz_spectral_1998,done_intrinsic_2012,petrucci_multiwavelength_2013,
petrucci_testing_2018,middei_high-energy_2019}.
Compared with the hot corona that is responsible for the hard X-ray continuum emission, the warm corona has larger 
optical depth $\sim10-20$ and lower temperature. On the other hand, the ionized reflection model
could equally fit the energy spectrum of the soft excess 
\citep[e.g.][]{ballantyne_evidence_2001,fabian_how_2002,crummy_explanation_2006}. In this scenario, the soft X-ray
emission lines in the reflection spectrum are emerging close to the black hole, and are thus greatly blurred by the 
strong gravitational field, leading to the soft excess.
One argument against the reflection scenario is that the strength of the soft excess component is uncorrelated with
the reflection strength \citep{boissay_hard_2016}.

In Fig.~\ref{fig:selfirr_enspec}, we see that for compact coronae with radius $\lesssim 4~\rm GM/c^2$, below $\sim 1 ~\rm keV$
the self-irradiation is a dominating source in interacting with the disc atmosphere.
By analyzing simultaneous UV/optical spectra of a subsample of CHEESE, we found that most of the AGNs in the subsample
have corona radius smaller than $4~\rm GM/c^2$ if assuming the height of the corona to be $10~\rm GM/c^2$ \citep{ursini_estimating_2020}.
Therefore, the self-irradiation is more relevant than the hot corona emission for the reflection features in the soft X-ray band,
including the soft excess. However, this component
has not been addressed in earlier works. On the other hand, given the energy of the high-energy cutoff of the self-irradiation component,
the self-irradiation has negligible effect on the reflection continuum or the iron K$\alpha$ line. Therefore, we do
not expect the strength of the soft excess to be correlated with the strength of the reflection emission.


\subsection{Anisotropic corona emission}
In \citet{zhang_constraining_2019}, we studied the angular dependence of Comptonized radiation. We found that for
disc-corona systems in AGNs, the observer at lower inclination sees harder and less luminous Comptonized radiation,
due to anisotropic seed photons. To assess the effect on the reflection spectrum, in \citet{zhang_constraining_2019}
we considered an ideal case where a spherical plasma cloud Comptonizes low-frequency unidirectional seed photons.
We found that the Comptonized radiation with inclination greater than $90^\circ$, which is an analog to corona emission
that reaches the accretion disc, is more luminous than the Comptonzied radiation with inclination less than $90^\circ$.

Here, we calculate the angular dependent Comptonized spectra in \textit{the rest frame of the extended corona}.
We perform the calculation for compact spherical coronae (with radii of $1~\rm GM/c^2$), for which we can safely
assume that the lensing and redshift properties of the corona emission is close to a lamp-post corona at the same
height. To assess these two effects, we sample $N_{\rm tot}$ photons originating from an isotropic
lamp-post corona, and count the number of photons that reach an inclination bin at infinity or a ring on the accretion
disc. The solid angle subtended by the inclination bin or the ring $d\Omega =
4\pi N/N_{\rm tot}$, where $N$ is the number of photons that falling into the inclination bin or the ring.
Having known $d\Omega$ and $g$, the spectral luminosity in the corona rest frame is
\begin{equation}
 L_E = \frac{4\pi \sum_i w_i E_{i,\infty}/g_i}{d\Omega \Delta E},
\end{equation}
where the sum is over all superphotons with energy between $E-\Delta E/2$ and $E+\Delta E/2$.

\begin{figure}
 \includegraphics[width=\columnwidth]{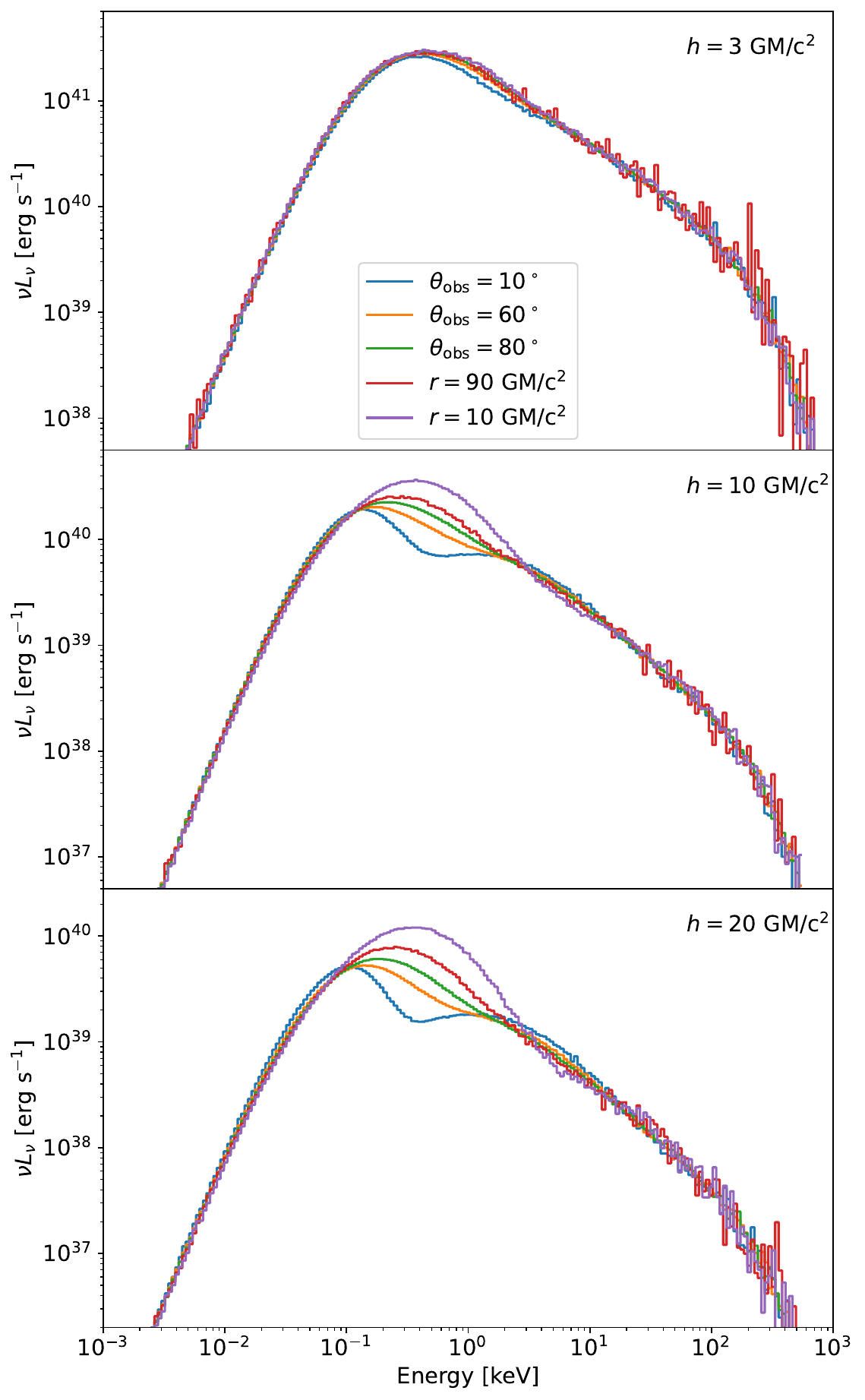}
 \caption{The energy spectra of the Comptonized emission, as seen by observers at various inclinations and by observers located
 at different radii on the accretion disc. The results for coronal heights of 3, 10, and 20 $\rm GM/c^2$ are plotted in the
 top, middle, and bottom panels, respectively.\label{fig:sphere_incl}}
\end{figure}

The spectra are presented in Fig.~\ref{fig:sphere_incl}, for corona heights of $3$, $10$, and $20~\rm GM/c^2$, respectively.
The dependence of the spectra on the inclination is similar with what we found in \citet{zhang_constraining_2019}.
This demonstrates our conclusion in \citet{zhang_constraining_2019} that the Comptonized radiation irradiating
the disc is less luminous and harder than that reaching infinity, as the radius decreases.
The spectra are also more anisotropic as the height increases, as the seed photons are getting more anisotropic.

\section{Summary}
\label{sec:summary}
In this work, we present the illumination and neutral iron K$\alpha$ emissivity profiles of accretion discs irradiated by extended coronae in simple spherical, lamp-post, and slab geometry. With our GR Monte Carlo radiative transfer code \textsc{Monk}, we calculated the energy spectrum of the illuminating radiation
in the rest frame of the accretion disc. As a result, we are able to calculate both the illumination and emissivity profiles while in most previous studies the authors did not make a clear distinction between the two.
For AGN discs, the emissivity profiles are in general steeper than the illumination
profiles for the parameters taken in the calculations; whereas for accretion discs in black hole X-ray binaries, the distinction is more
dramatic: the emissivity even decreases with decreasing radius in the innermost region of the disc.
We find out that the different behavior in AGNs and black hole X-ray binaries are due to the difference
in energy spectra of the illuminating radiation as seen by the disc.
This suggests that the emissivity profile of the iron K$\alpha$ line 
cannot be determined by black hole spin and corona geometry alone, but the energy spectrum of the illuminating radiation
has to be taken into account.
We study the dependence of the emissivity profile on the geometry of the corona and find the emissivity in the innermost region
of the disc steepens as the corona becomes less extended or as the corona has a larger height.

We also examine the effect of the self-irradiation. We find that for BHXRBs, the self-irradiation might be more important than the corona emission for the production of iron K$\alpha$ emission, especially when the corona has large extension and when the inner edge of the disc is close to the black hole. The self-irradiation in AGNs has lower temperature than the iron K edge, therefore having no
effect on the iron K emission. However, the self-irradiation may be important for the soft excess component.

\section*{acknowledgements}
We thank the anonymous referee for his/her useful comments that greatly improved the manuscript. WZ acknowledges the support by the Strategic Pioneer Program on Space Science, Chinese Academy of Sciences through grants XDA15310000, and the support by the Strategic Priority Research Program of the Chinese Academy of Sciences, Grant No. XDB0550200. This work is also supported by the National Natural Science Foundation of China
(grant 12333004). MD, MB, VK, and JS thank for the support from the GACR project 21-06825X and the institutional support from RVO:67985815.
This research makes uses of \textsc{Matplotlib} \citep{hunter_matplotlib:_2007},
a Python 2D plotting library which produces publication quality figures.

\section*{data availability}
The authors agree to make simulation data supporting the results in this paper available upon reasonable request.

\bibliographystyle{aasjournal}
\bibliography{/home/wdzhang/data/sim5corona/papers/emis_paper/v5/emis}

\begin{thebibliography}{}
\expandafter\ifx\csname natexlab\endcsname\relax\def\natexlab#1{#1}\fi
\providecommand{\url}[1]{\href{#1}{#1}}
\providecommand{\dodoi}[1]{doi:~\href{http://doi.org/#1}{\nolinkurl{#1}}}
\providecommand{\doeprint}[1]{\href{http://ascl.net/#1}{\nolinkurl{http://ascl.net/#1}}}
\providecommand{\doarXiv}[1]{\href{https://arxiv.org/abs/#1}{\nolinkurl{https://arxiv.org/abs/#1}}}

\bibitem[{{Ag{\'i}s-Gonz{\'a}lez} {et~al.}(2014){Ag{\'i}s-Gonz{\'a}lez},
  Miniutti, Kara, Fabian, Sanfrutos, Risaliti, Bianchi, Strotjohann, Saxton, \&
  Parker}]{agis-gonzalez_black_2014}
{Ag{\'i}s-Gonz{\'a}lez}, B., Miniutti, G., Kara, E., {et~al.} 2014, MNRAS, 443,
  2862, \dodoi{10.1093/mnras/stu1358}

\bibitem[{Arnaud {et~al.}(1985)Arnaud, {Branduardi-Raymont}, Culhane, Fabian,
  Hazard, McGlynn, Shafer, Tennant, \& Ward}]{arnaud_exosat_1985}
Arnaud, K.~A., {Branduardi-Raymont}, G., Culhane, J.~L., {et~al.} 1985, MNRAS,
  217, 105, \dodoi{10.1093/mnras/217.1.105}

\bibitem[{Ballantyne {et~al.}(2001)Ballantyne, Iwasawa, \&
  Fabian}]{ballantyne_evidence_2001}
Ballantyne, D.~R., Iwasawa, K., \& Fabian, A.~C. 2001, MNRAS, 323, 506,
  \dodoi{10.1046/j.1365-8711.2001.04234.x}

\bibitem[{Bianchi {et~al.}(2009)Bianchi, Guainazzi, Matt, Fonseca~Bonilla, \&
  Ponti}]{bianchi_caixa:_2009-1}
Bianchi, S., Guainazzi, M., Matt, G., Fonseca~Bonilla, N., \& Ponti, G. 2009,
  A\&A, 495, 421, \dodoi{10.1051/0004-6361:200810620}

\bibitem[{Blum {et~al.}(2009)Blum, Miller, Fabian, Miller, Homan, {van der
  Klis}, Cackett, \& Reis}]{blum_measuring_2009}
Blum, J.~L., Miller, J.~M., Fabian, A.~C., {et~al.} 2009, ApJ, 706, 60,
  \dodoi{10.1088/0004-637X/706/1/60}

\bibitem[{Boissay {et~al.}(2016)Boissay, Ricci, \& Paltani}]{boissay_hard_2016}
Boissay, R., Ricci, C., \& Paltani, S. 2016, A\&A, 588, A70,
  \dodoi{10.1051/0004-6361/201526982}

\bibitem[{Brenneman \& Reynolds(2006)}]{brenneman_constraining_2006}
Brenneman, L.~W., \& Reynolds, C.~S. 2006, ApJ, 652, 1028,
  \dodoi{10.1086/508146}

\bibitem[{Brenneman {et~al.}(2011)Brenneman, Reynolds, Nowak, Reis, Trippe,
  Fabian, Iwasawa, Lee, Miller, Mushotzky, Nandra, \&
  Volonteri}]{brenneman_spin_2011}
Brenneman, L.~W., Reynolds, C.~S., Nowak, M.~A., {et~al.} 2011, ApJ, 736, 103,
  \dodoi{10.1088/0004-637X/736/2/103}

\bibitem[{Chainakun \& Young(2017)}]{chainakun_investigating_2017}
Chainakun, P., \& Young, A.~J. 2017, MNRAS, 465, 3965,
  \dodoi{10.1093/mnras/stw2964}

\bibitem[{Crummy {et~al.}(2006)Crummy, Fabian, Gallo, \&
  Ross}]{crummy_explanation_2006}
Crummy, J., Fabian, A.~C., Gallo, L., \& Ross, R.~R. 2006, MNRAS, 365, 1067,
  \dodoi{10.1111/j.1365-2966.2005.09844.x}

\bibitem[{Czerny \& Elvis(1987)}]{czerny_constraints_1987}
Czerny, B., \& Elvis, M. 1987, ApJ, 321, 305, \dodoi{10.1086/165630}

\bibitem[{Dauser {et~al.}(2022)Dauser, Garc{\'i}a, Joyce, Licklederer, Connors,
  Ingram, Reynolds, \& Wilms}]{dauser_effect_2022}
Dauser, T., Garc{\'i}a, J.~A., Joyce, A., {et~al.} 2022, MNRAS, 514, 3965,
  \dodoi{10.1093/mnras/stac1593}

\bibitem[{Done {et~al.}(2012)Done, Davis, Jin, Blaes, \&
  Ward}]{done_intrinsic_2012}
Done, C., Davis, S.~W., Jin, C., Blaes, O., \& Ward, M. 2012, MNRAS, 420, 1848,
  \dodoi{10.1111/j.1365-2966.2011.19779.x}

\bibitem[{Dovciak {et~al.}(2014)Dovciak, Svoboda, Goosmann, Karas, Matt, \&
  Sochora}]{dovciak_xspec_2014}
Dovciak, M., Svoboda, J., Goosmann, R.~W., {et~al.} 2014, arXiv:1412.8627.
\newblock \doeprint{1412.8627}

\bibitem[{{El-Batal} {et~al.}(2016){El-Batal}, Miller, Reynolds, Boggs,
  Chistensen, Craig, Fuerst, Hailey, Harrison, Stern, Tomsick, Walton, \&
  Zhang}]{el-batal_nustar_2016}
{El-Batal}, A.~M., Miller, J.~M., Reynolds, M.~T., {et~al.} 2016, ApJL, 826,
  L12, \dodoi{10.3847/2041-8205/826/1/L12}

\bibitem[{Emmanoulopoulos {et~al.}(2011)Emmanoulopoulos, Papadakis, McHardy,
  Nicastro, Bianchi, \& Ar{\'e}valo}]{emmanoulopoulos_xmm-newton_2011}
Emmanoulopoulos, D., Papadakis, I.~E., McHardy, I.~M., {et~al.} 2011, MNRAS,
  415, 1895, \dodoi{10.1111/j.1365-2966.2011.18834.x}

\bibitem[{Esin {et~al.}(1997)Esin, McClintock, \&
  Narayan}]{esin_advection-dominated_1997}
Esin, A.~A., McClintock, J.~E., \& Narayan, R. 1997, ApJ, 489, 865

\bibitem[{Fabian {et~al.}(2002)Fabian, Ballantyne, Merloni, Vaughan, Iwasawa,
  \& Boller}]{fabian_how_2002}
Fabian, A.~C., Ballantyne, D.~R., Merloni, A., {et~al.} 2002, MNRAS, 331, L35,
  \dodoi{10.1046/j.1365-8711.2002.05419.x}

\bibitem[{Fabian {et~al.}(1989)Fabian, Rees, Stella, \&
  White}]{fabian_x-ray_1989}
Fabian, A.~C., Rees, M.~J., Stella, L., \& White, N.~E. 1989, MNRAS, 238, 729

\bibitem[{Fabian \& Ross(2010)}]{fabian_x-ray_2010}
Fabian, A.~C., \& Ross, R.~R. 2010, Space Science Reviews, 157, 167,
  \dodoi{10.1007/s11214-010-9699-y}

\bibitem[{Fabian {et~al.}(2009)Fabian, Zoghbi, Ross, Uttley, Gallo, Brandt,
  Blustin, Boller, {Caballero-Garcia}, Larsson, Miller, Miniutti, Ponti, Reis,
  Reynolds, Tanaka, \& Young}]{fabian_broad_2009}
Fabian, A.~C., Zoghbi, A., Ross, R.~R., {et~al.} 2009, Nature, 459, 540,
  \dodoi{10.1038/nature08007}

\bibitem[{George \& Fabian(1991)}]{george_x-ray_1991}
George, I.~M., \& Fabian, A.~C. 1991, MNRAS, 249, 352,
  \dodoi{10.1093/mnras/249.2.352}

\bibitem[{Gonzalez {et~al.}(2017)Gonzalez, Wilkins, \&
  Gallo}]{gonzalez_probing_2017}
Gonzalez, A.~G., Wilkins, D.~R., \& Gallo, L.~C. 2017, MNRAS, 472, 1932,
  \dodoi{10.1093/mnras/stx2080}

\bibitem[{Grevesse \& Sauval(1998)}]{grevesse_standard_1998}
Grevesse, N., \& Sauval, A.~J. 1998, Space Science Reviews, 85, 161,
  \dodoi{10.1023/A:1005161325181}

\bibitem[{Hunter(2007)}]{hunter_matplotlib:_2007}
Hunter, J.~D. 2007, Computing in Science Engineering, 9, 90,
  \dodoi{10.1109/MCSE.2007.55}

\bibitem[{Jiang {et~al.}(2022)Jiang, Dauser, Fabian, Alston, Gallo, Parker, \&
  Reynolds}]{jiang_xmm-newton_2022}
Jiang, J., Dauser, T., Fabian, A.~C., {et~al.} 2022, MNRAS, 514, 1107,
  \dodoi{10.1093/mnras/stac1144}

\bibitem[{Jiang {et~al.}(2019)Jiang, Walton, Fabian, \&
  Parker}]{jiang_relativistic_2019}
Jiang, J., Walton, D.~J., Fabian, A.~C., \& Parker, M.~L. 2019, MNRAS, 483,
  2958, \dodoi{10.1093/mnras/sty3228}

\bibitem[{Laor(1991)}]{laor_line_1991}
Laor, A. 1991, ApJ, 376, 90, \dodoi{10.1086/170257}

\bibitem[{Lucchini {et~al.}(2023)Lucchini, Mastroserio, Wang, Kara, Ingram,
  Garcia, Dauser, {van der Klis}, K{\"o}nig, Lewin, Nathan, \&
  Panagiotou}]{lucchini_investigating_2023}
Lucchini, M., Mastroserio, G., Wang, J., {et~al.} 2023, ApJ, 951, 19,
  \dodoi{10.3847/1538-4357/acd24f}

\bibitem[{Magdziarz {et~al.}(1998)Magdziarz, Blaes, Zdziarski, Johnson, \&
  Smith}]{magdziarz_spectral_1998}
Magdziarz, P., Blaes, O.~M., Zdziarski, A.~A., Johnson, W.~N., \& Smith, D.~A.
  1998, MNRAS, 301, 179, \dodoi{10.1046/j.1365-8711.1998.02015.x}

\bibitem[{Martocchia {et~al.}(2000)Martocchia, Karas, \&
  Matt}]{martocchia_effects_2000}
Martocchia, A., Karas, V., \& Matt, G. 2000, MNRAS, 312, 817,
  \dodoi{10.1046/j.1365-8711.2000.03205.x}

\bibitem[{Middei {et~al.}(2019)Middei, Bianchi, Petrucci, Ursini, Cappi,
  De~Marco, De~Rosa, Malzac, Marinucci, Matt, Ponti, \&
  Tortosa}]{middei_high-energy_2019}
Middei, R., Bianchi, S., Petrucci, P.-O., {et~al.} 2019, MNRAS, 483, 4695,
  \dodoi{10.1093/mnras/sty3379}

\bibitem[{Miller(2007)}]{miller_relativistic_2007}
Miller, J.~M. 2007, ARA\&A, 45, 441,
  \dodoi{10.1146/annurev.astro.45.051806.110555}

\bibitem[{Miller {et~al.}(2009)Miller, Reynolds, Fabian, Miniutti, \&
  Gallo}]{miller_stellar-mass_2009}
Miller, J.~M., Reynolds, C.~S., Fabian, A.~C., Miniutti, G., \& Gallo, L.~C.
  2009, ApJ, 697, 900, \dodoi{10.1088/0004-637X/697/1/900}

\bibitem[{Miller {et~al.}(2013)Miller, Parker, Fuerst, Bachetti, Harrison,
  Barret, Boggs, Chakrabarty, Christensen, Craig, Fabian, Grefenstette, Hailey,
  King, Stern, Tomsick, Walton, \& Zhang}]{miller_nustar_2013}
Miller, J.~M., Parker, M.~L., Fuerst, F., {et~al.} 2013, ApJL, 775, L45,
  \dodoi{10.1088/2041-8205/775/2/L45}

\bibitem[{Miller {et~al.}(2018)Miller, Gendreau, Ludlam, Fabian, Altamirano,
  Arzoumanian, Bult, Cackett, Homan, Kara, Neilsen, Remillard, \&
  Tombesi}]{miller_nicer_2018}
Miller, J.~M., Gendreau, K., Ludlam, R.~M., {et~al.} 2018, ApJL, 860, L28,
  \dodoi{10.3847/2041-8213/aacc61}

\bibitem[{Miniutti {et~al.}(2003)Miniutti, Fabian, Goyder, \&
  Lasenby}]{miniutti_lack_2003}
Miniutti, G., Fabian, A.~C., Goyder, R., \& Lasenby, A.~N. 2003, MNRAS, 344,
  L22, \dodoi{10.1046/j.1365-8711.2003.06988.x}

\bibitem[{Miniutti {et~al.}(2009{\natexlab{a}})Miniutti, Panessa, {de Rosa},
  Fabian, Malizia, Molina, Miller, \& Vaughan}]{miniutti_intermediate_2009}
Miniutti, G., Panessa, F., {de Rosa}, A., {et~al.} 2009{\natexlab{a}}, MNRAS,
  398, 255, \dodoi{10.1111/j.1365-2966.2009.15092.x}

\bibitem[{Miniutti {et~al.}(2009{\natexlab{b}})Miniutti, Ponti, Greene, Ho,
  Fabian, \& Iwasawa}]{miniutti_xmm_2009}
Miniutti, G., Ponti, G., Greene, J.~E., {et~al.} 2009{\natexlab{b}}, MNRAS,
  394, 443, \dodoi{10.1111/j.1365-2966.2008.14334.x}

\bibitem[{Page {et~al.}(2004)Page, Schartel, Turner, \&
  O'Brien}]{page_xmm-newton_2004}
Page, K.~L., Schartel, N., Turner, M. J.~L., \& O'Brien, P.~T. 2004, MNRAS,
  352, 523, \dodoi{10.1111/j.1365-2966.2004.07939.x}

\bibitem[{Parker {et~al.}(2015)Parker, Tomsick, Miller, Yamaoka, Lohfink,
  Nowak, Fabian, Alston, Boggs, Christensen, Craig, F{\"u}rst, Gandhi,
  Grefenstette, Grinberg, Hailey, Harrison, Kara, King, Stern, Walton, Wilms,
  \& Zhang}]{parker_nustar_2015}
Parker, M.~L., Tomsick, J.~A., Miller, J.~M., {et~al.} 2015, ApJ, 808, 9,
  \dodoi{10.1088/0004-637X/808/1/9}

\bibitem[{Petrucci {et~al.}(2018)Petrucci, Ursini, De~Rosa, Bianchi, Cappi,
  Matt, Dadina, \& Malzac}]{petrucci_testing_2018}
Petrucci, P.-O., Ursini, F., De~Rosa, A., {et~al.} 2018, A\&A, 611, A59,
  \dodoi{10.1051/0004-6361/201731580}

\bibitem[{Petrucci {et~al.}(2013)Petrucci, Paltani, Malzac, Kaastra, Cappi,
  Ponti, De~Marco, Kriss, Steenbrugge, Bianchi, {Branduardi-Raymont},
  Mehdipour, Costantini, Dadina, \&
  Lubi{\'n}ski}]{petrucci_multiwavelength_2013}
Petrucci, P.-O., Paltani, S., Malzac, J., {et~al.} 2013, A\&A, 549, A73,
  \dodoi{10.1051/0004-6361/201219956}

\bibitem[{Porquet {et~al.}(2004)Porquet, Reeves, O'Brien, \&
  Brinkmann}]{porquet_xmm-newton_2004}
Porquet, D., Reeves, J.~N., O'Brien, P., \& Brinkmann, W. 2004, A\&A, 422, 85,
  \dodoi{10.1051/0004-6361:20047108}

\bibitem[{Pounds {et~al.}(1986)Pounds, Warwick, Culhane, \& {de
  Korte}}]{pounds_soft_1986}
Pounds, K.~A., Warwick, R.~S., Culhane, J.~L., \& {de Korte}, P. a.~J. 1986,
  MNRAS, 218, 685, \dodoi{10.1093/mnras/218.4.685}

\bibitem[{Reynolds {et~al.}(2012)Reynolds, Brenneman, Lohfink, Trippe, Miller,
  Fabian, \& Nowak}]{reynolds_monte_2012}
Reynolds, C.~S., Brenneman, L.~W., Lohfink, A.~M., {et~al.} 2012, ApJ, 755, 88,
  \dodoi{10.1088/0004-637X/755/2/88}

\bibitem[{Ricci {et~al.}(2014)Ricci, Tazaki, Ueda, Paltani, Boissay, \&
  Terashima}]{ricci_suzaku_2014}
Ricci, C., Tazaki, F., Ueda, Y., {et~al.} 2014, ApJ, 795, 147,
  \dodoi{10.1088/0004-637X/795/2/147}

\bibitem[{Risaliti {et~al.}(2013)Risaliti, Harrison, Madsen, Walton, Boggs,
  Christensen, Craig, Grefenstette, Hailey, Nardini, Stern, \&
  Zhang}]{risaliti_rapidly_2013}
Risaliti, G., Harrison, F.~A., Madsen, K.~K., {et~al.} 2013, Nature, 494, 449,
  \dodoi{10.1038/nature11938}

\bibitem[{Schmoll {et~al.}(2009)Schmoll, Miller, Volonteri, Cackett, Reynolds,
  Fabian, Brenneman, Miniutti, \& Gallo}]{schmoll_constraining_2009}
Schmoll, S., Miller, J.~M., Volonteri, M., {et~al.} 2009, ApJ, 703, 2171,
  \dodoi{10.1088/0004-637X/703/2/2171}

\bibitem[{Singh {et~al.}(1985)Singh, Garmire, \&
  Nousek}]{singh_observations_1985}
Singh, K.~P., Garmire, G.~P., \& Nousek, J. 1985, ApJ, 297, 633,
  \dodoi{10.1086/163560}

\bibitem[{Sun {et~al.}(2018)Sun, Guainazzi, Ni, Wang, Qian, Shi, Wang, \&
  Bambi}]{sun_multi-epoch_2018}
Sun, S., Guainazzi, M., Ni, Q., {et~al.} 2018, MNRAS, 478, 1900,
  \dodoi{10.1093/mnras/sty1233}

\bibitem[{Svoboda {et~al.}(2012)Svoboda, Dov{\v c}iak, Goosmann, Jethwa, Karas,
  Miniutti, \& Guainazzi}]{svoboda_origin_2012}
Svoboda, J., Dov{\v c}iak, M., Goosmann, R.~W., {et~al.} 2012, A\&A, 545, A106,
  \dodoi{10.1051/0004-6361/201219701}

\bibitem[{Tan {et~al.}(2012)Tan, Wang, Shu, \& Zhou}]{tan_possible_2012}
Tan, Y., Wang, J.~X., Shu, X.~W., \& Zhou, Y. 2012, ApJL, 747, L11,
  \dodoi{10.1088/2041-8205/747/1/L11}

\bibitem[{Taylor \& Reynolds(2018)}]{taylor_exploring_2018}
Taylor, C., \& Reynolds, C.~S. 2018, ApJ, 855, 120,
  \dodoi{10.3847/1538-4357/aaad63}

\bibitem[{Turner \& Pounds(1988)}]{turner_variability_1988}
Turner, T.~J., \& Pounds, K.~A. 1988, MNRAS, 232, 463,
  \dodoi{10.1093/mnras/232.2.463}

\bibitem[{Ursini {et~al.}(2020)Ursini, Dov{\v c}iak, Zhang, Matt, Petrucci, \&
  Done}]{ursini_estimating_2020}
Ursini, F., Dov{\v c}iak, M., Zhang, W., {et~al.} 2020, A\&A, 644, A132,
  \dodoi{10.1051/0004-6361/202039158}

\bibitem[{Verner \& Yakovlev(1995)}]{verner_analytic_1995}
Verner, D.~A., \& Yakovlev, D.~G. 1995, Astronomy Astrophysics Supplement
  Series, 109, 125

\bibitem[{Walter \& Fink(1993)}]{walter_ultraviolet_1993}
Walter, R., \& Fink, H.~H. 1993, A\&A, 274, 105

\bibitem[{Wilkins \& Fabian(2012)}]{wilkins_understanding_2012}
Wilkins, D.~R., \& Fabian, A.~C. 2012, MNRAS, 424, 1284,
  \dodoi{10.1111/j.1365-2966.2012.21308.x}

\bibitem[{Wilkins {et~al.}(2020)Wilkins, Garc{\'i}a, Dauser, \&
  Fabian}]{wilkins_returning_2020}
Wilkins, D.~R., Garc{\'i}a, J.~A., Dauser, T., \& Fabian, A.~C. 2020, MNRAS,
  498, 3302, \dodoi{10.1093/mnras/staa2566}

\bibitem[{Xu {et~al.}(2018{\natexlab{a}})Xu, Harrison, Garc{\'i}a, Fabian,
  F{\"u}rst, Gandhi, Grefenstette, Madsen, Miller, Parker, Tomsick, \&
  Walton}]{xu_reflection_2018}
Xu, Y., Harrison, F.~A., Garc{\'i}a, J.~A., {et~al.} 2018{\natexlab{a}}, ApJL,
  852, L34, \dodoi{10.3847/2041-8213/aaa4b2}

\bibitem[{Xu {et~al.}(2018{\natexlab{b}})Xu, Harrison, Kennea, Walton, Tomsick,
  Miller, Barret, Fabian, Forster, F{\"u}rst, Gandhi, \&
  Garc{\'i}a}]{xu_hard_2018}
Xu, Y., Harrison, F.~A., Kennea, J.~A., {et~al.} 2018{\natexlab{b}}, ApJ, 865,
  18, \dodoi{10.3847/1538-4357/aada03}

\bibitem[{Zhang {et~al.}(2019)Zhang, Dov{\v c}iak, \&
  Bursa}]{zhang_constraining_2019}
Zhang, W., Dov{\v c}iak, M., \& Bursa, M. 2019, ApJ, 875, 148,
  \dodoi{10.3847/1538-4357/ab1261}

\end{thebibliography}

\label{lastpage}
\end{document}